# Through Skull Fluorescence Imaging of the Brain in a New Near-Infrared Window

Guosong Hong[1,5], Shuo Diao[1,5], Junlei Chang[2,5], Alexander L. Antaris[1], Changxin Chen[1], Bo Zhang[1], Su Zhao[1], Dmitriy N. Atochin[3], Paul L. Huang[3], Katrin I. Andreasson[4], Calvin J. Kuo[2] and Hongjie Dai[1]

[1] Department of Chemistry, Stanford University, Stanford, California 94305, USA.

[2] Division of Hematology, School of Medicine, Stanford University, Stanford, California 94305, USA.

[3] Cardiovascular Research Center and Cardiology Division, Massachusetts General Hospital, Harvard Medical School, Charlestown, Massachusetts 02129, USA.

[4] Department of Neurology and Neurological Sciences, Stanford University Medical Center, Stanford, California 94305, USA.

[5] These authors contribute equally to this work.

Correspondence and requests for materials should be addressed to H.D. (email: hdai@stanford.edu) or C.J.K. (email: cjkuo@stanford.edu).

## Abstract

To date, brain imaging has largely relied on X-ray computed tomography and magnetic resonance angiography with limited spatial resolution and long scanning times. Fluorescence-based brain imaging in the visible and traditional near-infrared regions (400-900 nm) is an alternative but currently requires craniotomy, cranial windows and skull thinning techniques, and the penetration depth is limited to 1-2 mm due to light scattering. Here, we report through-scalp and through-skull fluorescence imaging of mouse cerebral vasculature without craniotomy utilizing the intrinsic photoluminescence of single-walled carbon nanotubes in the 1.3-1.4 micrometre near-infrared window (NIR-IIa window). Reduced photon scattering in this spectral region allows fluorescence imaging reaching a depth of >2 mm in mouse brain with sub-10 micrometre resolution. An imaging rate of ~5.3 frames/s allows for dynamic recording of blood perfusion in the cerebral vessels with sufficient temporal resolution, providing real-time assessment of blood flow anomaly in a mouse middle cerebral artery occlusion stroke model.

## Introduction

The essential functions of the brain dictate that any significant cerebral dysfunction or disease, such as the cerebrovascular disease, could lead to severe morbidity or mortality.[1] Brain imaging has relied on X-ray computed tomography (CT) and magnetic resonance angiography (MRA) to reach sufficient penetration depth,[2-4] but these modalities are limited in spatial resolution (~ sub-millimeter) and long scanning times (~ minutes),[5-7] impairing the resolution of small vessels and dynamic blood flow in the brain. On the other hand, fluorescence-based optical imaging of the brain in the visible and traditional near-infrared regions (400-900 nm) has relied on craniotomy, cranial windows and skull thinning techniques[8-11] and the penetration depth for *in vivo* brain imaging has been limited to a depth of 1-2 mm[8,12-14] by light scattering.[15,16]

Recently, we and others have shown that biological imaging with carbon nanotube and quantum dot fluorescent agents in the long near-infrared region (1.0-1.7 μm, named NIR-II region) can benefit from reduced tissue scattering and autofluorescence, reaching deeper penetration depths *in vivo* than traditional NIR (NIR-I, 750-900 nm) imaging techniques.[17-29] Photon scattering scales as $\lambda^{-\alpha}$, where $\alpha = 0.2$ to $4$ for biological tissues. The reduced scattering in the NIR-II region has allowed *in vivo* fluorescence imaging of blood vessels in the hindlimb of mice with ~30 μm lateral resolution at > 1 mm depth.[22] Here, we report the first non-invasive brain imaging in a narrow 1.3-1.4 μm region (named as the NIR-IIa region), allowing penetration through intact scalp and skull and resolving cerebral vasculatures with a previously unattainable spatial resolution of sub-10 μm at a depth of >2 mm underneath the surface of the scalp skin in an epifluorescence imaging mode. The 1.3-1.4 μm NIR-IIa window is rationally chosen for *in vivo* imaging so as to minimize scattering by rejecting shorter wavelength photons than 1.3 μm while avoiding increase light absorption by water vibrational overtone modes above ~1.4 μm. Further, dynamic NIR-IIa cerebrovascular imaging with high temporal resolution (< 200 ms/frame) was used to reveal drastically reduced blood flow due to arterial occlusion in an acute stroke model on mice.

## Results

**Phantom Imaging in NIR-I, NIR-II and NIR-IIa Regions.** We first used a highly scattering medium (Intralipid ®, 1 wt% aqueous solution) as a mimic to biological tissue[20] and studied the penetration depths of fluorescence imaging in the NIR-I (<900 nm), NIR-II (1.0-1.7 μm) and NIR-IIa (1.3-1.4 μm) regions. High pressure CO conversion (HiPCO) single-walled carbon nanotube[30] (SWNT)-IRDye800 conjugates excited by an 808-nm laser were used as emitters fluorescing in a wide 800 – 1700 nm span (for side-by-side comparing imaging with IRDye800 that emits in the 800-900 nm NIR-I window, and with SWNT that emits in the 1000-1700 nm NIR-II window),[31,32] using optical filters to select detection and imaging wavelength ranges (**Fig. 1a**). When a capillary tube filled with the SWNT-IRDye800 solution was immersed in an Intralipid solution at 1 mm depth, sharp images were obtained in all three spectral windows (**Fig. 1b**, top). However, when the sample immersion depth increased to 1 cm in Intralipid, scattering of photons became obvious in the entire NIR-II (1.0-1.7 μm) region and was more severe in the NIR-I (< 900 nm) region, causing the image of the capillary tube to be smeared and invisible (**Fig. 1b,** bottom). In contrast, an image taken in the 1.3-1.4 μm NIR-IIa region resolved sharp edges of the capillary tube even at 1 cm immersion depth in Intralipid (**Fig. 1b** bottom right image), suggesting reduced scattering. The observed wavelength dependence of phantom imaging (**Fig. 1b** & **S1**) could be rationalized by considering the measured scattering coefficient vs. wavelength for 1% Intralipid (red curve, **Fig. 1c**, in agreement with previously reported reduced scattering coefficient of $\mu'_s = 1.6\,\lambda^{-2.4}$, blue curve. In this case, and those to follow, $\mu'_s$ is the scattering coefficient in $mm^{-1}$ and $\lambda$ is the wavelength in $\mu m$, with the prefactor in appropriate unit to make units consistent on both sides of the equation).[33] The curves exhibited photon scattering scaling inversely proportional with wavelength. Rejecting shorter wavelength photons below 1.3 $\mu m$ led to drastically improved imaging by avoiding feature smearing and background signals caused by scattering of short-wavelength photons. The increased absorption in the >1.4-$\mu m$ region (peak shown in **Fig. 1c**) was due to light excitation of the first overtone vibration of water molecules in the Intralipid solution (**Fig. 1c** and **S2**)[34] and was also excluded from the NIR-IIa imaging window.

**Live Mouse Brain Imaging in NIR-I, NIR-II and NIR-IIa Regions.** We performed non-invasive *in vivo* cerebrovascular fluorescence imaging of healthy C57Bl/6 mice with the same SWNT-IRDye800 conjugates in several spectral windows. A solution of SWNT-IRDye800 was injected intravenously through the tail (see Methods) of a mouse. The head of the mouse with hair shaved off (**Fig. 2a**) was imaged without craniotomy in the NIR-I (< 900 nm), NIR-II (1.0-1.7 $\mu m$) and NIR-IIa (1.3-1.4 $\mu m$) regions respectively (pixel size = 78 $\mu m$) under an 808-nm laser illumination (**Fig. S3**). In contrast to the blurry vasculatures imaged in the < 900 nm NIR-I region (**Fig. 2b**), much sharper images were obtained by detecting longer wavelength photons (**Fig. 2c**), and the sharpest and highest resolution images of brain vessels were seen in the 1.3-1.4 $\mu m$ NIR-IIa region (**Fig. 2d**). Similar to the Intralipid phantom imaging case, rejecting the shorter wavelength < 1.3 $\mu m$ photons significantly reduced background signals and vessel blurriness caused by scattering, and improved signal/background ratio and imaging quality. The cerebral angiogram taken in the 1.3-1.4 μm NIR-IIa region clearly showed the inferior cerebral vein (labeled as 1), the superior sagittal sinus (labeled as 2) and transverse sinus (labeled as 3) at ~1-2 mm depths under the scalp skin (**Fig. S4**), along with other cortical vessels in both cerebral hemispheres (**Fig. 2d**). A 360-degree rotational view of the mouse head was also recorded in the 1.3-1.4 μm NIR-IIa region to show a 3D perspective of the mouse head (**Movie S1**).

The wavelength dependence of fluorescence imaging of mouse brain can be gleaned by inspecting the extinction spectra of the scalp skin and the cranial bone (i.e., the skull), both of which feature a water overtone vibration absorption peak above ~ 1.4 μm and a declining photon scattering profile (**Fig. 2e**). To examine the effect of scattering, we plotted reduced scattering coefficients $\mu_s'$ of these tissues versus wavelength $\lambda$ (**Fig. 2f**), based on the empirical formulas given in previous literature ($\mu_s'$ (scalp) $= 0.11\,\lambda^{-4} + 1.61\,\lambda^{-0.22}$, where the two terms are attributed to Rayleigh and Mie scattering respectively; $\mu_s'$ (skull) $= 1.72\,\lambda^{-0.65}$; and $\mu_s'$ (brain tissue) $= 4.72\,\lambda^{-2.07}$) [8,35,36]. It can be seen from the plotted scattering profiles that the reduced photon scattering coefficients in the NIR-IIa range (1.54 mm$^{-1}$ for scalp and 1.42 mm$^{-1}$ for skull at 1350 nm) were lower than in the traditional NIR-I window (1.96 mm$^{-1}$ for scalp and 1.99 mm$^{-1}$ for skull at 800 nm). A simple estimate suggested that the lower scattering coefficients in NIR-IIa corresponded to ~47% fewer scattered photons through the scalp and skull than in the NIR-I region based on the measured scalp thickness of 0.7 mm and the measured skull thickness of 0.6 mm. These factors led to drastically improved imaging of mouse cortical vessels underneath the cranial bone in the NIR-IIa region (**Fig. 2d vs. 2b**).

**High Resolution Cerebral Imaging in the NIR-IIa Window.** We then used microscopic objectives to image cerebral vessels with high resolution at 15-fold (pixel size ~ 5 μm) and 30-fold (pixel size ~ 2.5 μm) higher magnification than used for **Fig. 2**. A home-made stereotactic platform was used to eliminate motion of the mouse head and a 3D translational stage with digital readout allowed us to measure the imaging focal depth inside the brain (**Fig. 3a&b**). Further, chemical separation of large-diameter semiconducting (LS) nanotubes in the HiPCO material (see Methods for separation details) was performed and the LS tubes with fluorescence emission biased towards the favorable 1.3-1.4 μm region were used for brain imaging to optimize signal on the per unit mass basis of the injected nanotube dose (**Fig. 3c** and **S5**). We imaged the entire mouse head (**Fig. 3d**), zoomed into its left hemisphere (**Fig. 3e**) and then performed microscopic imaging near the location of a cortical vessel branching from the superior sagittal sinus. A typical microscopic image of the brain vessels taken in the 1.3-1.4 μm NIR-IIa region focused at a depth of 2.6 mm below the surface of the scalp (**Fig. 3f**; depth of focal plane was determined by the axial travel distance of the digitally controlled stage relative to the objective, as described in Methods) revealed many tiny capillary vessels branching from the larger vessels. Cross-sectional intensity profile of one of the capillary vessels in **Fig. 3g** showed a Gaussian-fitted full width at half maximum (FWHM) of 6.6 μm (**Fig. 3g** inset). Based on our *ex vivo* measurement of the total thickness of scalp skin, skull and the meninges as ~1.3 mm (**Fig. S4**), this capillary vessel located at a depth of 2.6 mm was ~ 1.3 mm deep (2.6 mm-1.3 mm) within the brain tissue. This represented the highest resolution non-invasive fluorescence imaging of brain capillary vessels reported to date (see **Fig. 3h-k** and **Fig. S6** for more brain capillaries imaged with 3 different mice at 1-3 mm depths).

Statistical analysis of measured capillary vessel widths revealed an average vessel width of 9.4 ± 2.5 μm, based on 63 different capillary vessels with widths ranging from 5 μm to 15 μm taken from a series of microscopic cerebral vascular images (**Fig. S7**). The brain imaging depth was limited to < 3 mm for our wide-field epifluorescence microscopic imaging setup. This limit was likely set by the interference from out-of-focus signals, most notably fluorescence from foreground vessels on the path of the laser beam prior to the focal plane.[37] We envisage a two-photon fluorescence microscope with NIR-IIa excitation should solve this problem and generate optically sectioned images with three-dimensional reconstruction of capillary networks to even deeper penetration depths in the brain, using traditional fluorescent dyes excitable around 600-700 nm in the one-photon absorption process.

**Dynamic NIR-IIa Fluorescence Imaging of Cerebral Blood Perfusion.** Immediately after tail-vein injection of SWNTs into a healthy control mouse (Mouse C1), we performed dynamic brain imaging (frame rate = 5.3 frame per second) and observed the NIR-IIa signal arising from the lateral sulcus on both sides of the cerebrum within 3 s (**Fig. 4a** & **Movie S2**), before the interior cerebral vein, the superior sagittal sinus and the transverse sinus started showing up after 4 s owing to the outflow of SWNTs into the venous vessels (**Fig. 4b&c**). Principal component analysis (PCA) was applied to the time-course images over a time course of ~20 s and discriminated the arterial vessels (red, **Fig. 4d**) from the venous vessels (blue, **Fig. 4e**) based on their haemodynamic difference (**Fig. 4f**).[20,22] The injected SWNTs kept circulating in the blood for at least 3 h post injection, allowing for static imaging in this time window (**Fig. S8**). We then repeated the study in mice with surgically induced middle cerebral artery occlusion (MCAO) as a model for stroke to the left cerebral hemisphere. After injection of an SWNT solution into a mouse with MCAO (Mouse M1), the right hemisphere exhibited very similar blood flow to the healthy Mouse C1, while the left hemisphere with MCAO showed a marked delay of blood flow revealed by NIR-IIa fluorescence (**Fig. 4g-i, Movie S3**). PCA analysis of Mouse M1 revealed a more extended venous vessel network (blue) in the right cerebral hemisphere than in the left, and the arterial vessels (red) only showed up in the intact right hemisphere (**Fig. 4j-l**). Similar results were reproduced with several healthy mice (**Fig. S9a-h**) and MCAO diseased mice (**Fig. S10a-l**).

The dynamic NIR-IIa fluorescence imaging allowed us to quantify cerebral blood perfusion by measuring average NIR-IIa intensity in the region of MCAs.[22] In the healthy Mouse C1, the linearly rising fluorescence signals in both cerebral hemispheres had almost identical slopes, suggesting a relative perfusion of ~1 (see SI for perfusion analysis) (**Fig. 4m**; **Fig. S9i&j** for repeat). However, in Mouse M1 with MCAO, the relative perfusion in the occluded left hemisphere was only 0.159 (**Fig. 4n**; **Fig. S10m-o** for repeat). Another group of mice (*n*=4) with cerebral hypoperfusion as a model of circulatory shock was imaged (**Fig. S11**) to reveal a decrease of blood perfusion of ~20% compared to a dramatic decrease of blood perfusion of ~85% in the MCAO group (red bars, **Fig. 4o**). The data agreed with laser Doppler measurements (blue bars, **Fig. 4o**). One advantage of NIR-IIa dynamic fluorescence imaging is that it tracks blood flow in a wide-field imaging setup, which allows for simultaneous tracking of blood flow in multiple vessels and could potentially provide a complementary method to two-photon microscopy based blood flow measurement, which performs line scan along the direction of one single vessel at a time with kHz temporal resolution.[38-40]

## Discussion

The NIR-IIa fluorescence imaging method is advantageous for brain imaging in terms of high resolution, deep penetration depth and non-invasive nature without the need for craniotomy. Previously, imaging in the traditional NIR-I region (750-900 nm) has observed sub-10 μm capillary vessels at ~350 μm depth with craniotomy to remove scattering extracerebral tissues,[12] much shallower than the imaging depth of >2 mm below the scalp in the NIR-IIa region without craniotomy. This striking difference can be rationalized by considering the highly scattering nature of the brain tissue, with a large scattering coefficient of ~ 7.49 mm$^{-1}$ at 800 nm versus ~ 2.54 mm$^{-1}$ at 1350 nm (**Fig. 2f**), suggesting ~$10^4$ times greater scattered photons in NIR-I than in NIR-IIa through a ~ 2 mm thick brain tissue. Two- and multi-photon microscopy techniques have reported a typical penetration depth of 1-2 mm in the brain tissue, facilitated by highly localized non-linear excitation and detection;[8,13,14,41] but scalp and skull usually need to be removed and replaced with a cranial window,[8,9] or polished and thinned[10,11] to allow for deeper penetration depth. Similarly, the invasive removal of the overlying tissues such as the scalp and the skull is involved for brain angiography with laser speckle contrast imaging (LSCI)[42,43] and optical frequency domain imaging (OFDI),[44] which have been reported to reach a high spatial resolution of ~10 μm *in vivo*.[44,45] To avoid the invasive treatment of the extracerebral tissue, optical coherence tomography (OCT) and photoacoustic microscopy have both shown the capability of visualizing cortical vasculature at the cost of spatial resolution (10-100 μm), unable to resolve sub-10 μm capillary vessels *in vivo*. In light of the possible inflammatory responses such as altered pial blood vessels induced by invasive extracerebral treatment,[10] the NIR-IIa fluorescence based brain angiography provides a completely non-invasive method to visualize the brain vasculature in live mouse with unprecedented resolution and penetration depth.

Although the maximum penetration depth of < 3 mm in this work prevents it from clinical imaging of much deeper vasculature inside the human brain, our new brain imaging method based on the long-wavelength NIR-IIa fluorescence can serve as a non-invasive biomedical imaging tool of understanding the mechanisms of the impaired neurovascular and neurometabolic regulation on both cortical and sub-cortical levels in animal model studies. Besides the cerebrovascular dynamics probed with NIR-IIa fluorescence in this work, we also envisage non-invasive brain imaging of neuronal and/or glial network activity with cellular resolution in live mice using neuronal activity indicators labeled with NIR-IIa fluorophores. In the long term, we anticipate fluorescence imaging using novel, metabolizable fluorophores with even longer emission wavelengths and less scattering to be eventually applied to diagnosis and monitor of cerebral vascular anomaly such as arteriovenous malformation and brain tumor, and other neurovascular disregulations involving rapid modification of cerebral blood flow, such as migraine.

## Conclusions

In this work, we explored a new biological transparent sub-window in the 1.3-1.4 μm, i.e., NIR-IIa region and performed non-invasive brain imaging in this window by penetrating through the intact skin and skull. We resolved cerebral vasculatures with a high spatial resolution of sub-10 μm at a depth of >2 mm in an epifluorescence imaging mode. Compare to previous NIR-II work, we found that the 1.3-1.4 μm NIR-IIa window for *in vivo* imaging can further reduce tissue scattering by rejecting shorter wavelength photons than 1.3 μm. The truly non-invasive nature and dynamic imaging capability of NIR-IIa cerebrovascular imaging could allow for high spatial and temporal resolution imaging to follow biological processes in the brain at the molecular scale. Future work will develop novel high quantum yield NIR-IIa fluorophores, 3D *in vivo* imaging through confocal or two photon techniques, all of which could lead to further advances of *in vivo* optical imaging.

## Methods

**Preparation of water-soluble and biocompatible SWNT-IRDye800 multicolor emitter.** The preparation of water soluble and biocompatible SWNTs is based upon the detailed procedure described in another publication of our group with some modifications.[46] In brief, raw HiPCO SWNTs (Unidym) were suspended in an aqueous solution containing 1 wt% sodium deoxycholate by 1 hour of bath sonication. This suspension was then ultracentrifuged at 300,000 g to remove the nanotube bundles and other large aggregates. The supernatant was retained after ultracentrifugation and 0.75 mg ml$^{-1}$ of DSPE-mPEG(5 kDa) (1,2-distearoyl-*sn*-glycero-3-phosphoethanolamine-N-[methoxy(polyethyleneglycol, 5000)], Laysan Bio) along with 0.25 mg ml$^{-1}$ of DSPE-PEG(5 kDa)-NH$_2$ (1,2-distearoyl-*sn*-glycero-3-phosphoethanolamine-N-[amino(polyethyleneglycol, 5000)], Laysan Bio) was added to the supernatant. The resulting suspension was sonicated in a bath sonicator briefly for 5 min and then dialyzed in a 3500 Da membrane (Fisher) against 1x phosphate buffer saline (PBS) solution at pH 7.4 with a minimum of six bath changes and a minimum of two hours between changes of water bath. To remove aggregates formed during dialysis, the suspension was ultracentrifuged again for 30 min at 300,000 g. This surfactant-exchanged SWNT sample has lengths ranging between 100 nm and 2.0 μm with the average length of ~500 nm. These amine-functionalized SWNTs were further conjugated with IRDye800 dye molecules according to the protocol developed in our group.[47] Briefly, the as-made SWNT solution was concentrated down to ~2 μM after removal of excess surfactant through 30-kDa centrifugal filters (Amicon) and was mixed with 0.1 mM IRDye800 NHS ester (LI-COR) dissolved in dimethylsulfoxide (DMSO). The reaction was allowed to proceed for 1 h at pH 7.4 before purification to remove excess IRDye800 by filtration through the 30-kDa filters. The as-made SWNT-IRDye800 conjugate solution was kept at 4 ℃ in the refrigerator and away from light to avoid photobleaching of the IRDye800 fluorescence.

**Sorting of large-diameter semiconducting (LS) nanotubes.** A 1-cm diameter column filled with 20 ml of ally dextran-based size-exclusion gel (Sephacryl S-200, GE Healthcare) was used as filtration medium to perform diameter separation of HiPCO SWNTs.[48] 2.5 mg of raw HiPCO SWNTs (Unidym) was sonicated in water with 1 wt% sodium cholate (SC) for 1 h, followed by an ultracentrifugation at 300,000 g for 30 min to remove large bundles and aggregates. 80% of the SC-dispersed supernatant was collected and diluted with same volume of 1 wt% sodium dodecyl sulfate (SDS) solution to make a mixture of 0.5 wt% SC and 0.5 wt% SDS surfactants, before adding to the gel column. Metallic SWNTs passed the column directly, while semiconducting SWNTs were trapped. Finally, a 0.6 wt% SC/0.4 wt% SDS mixed surfactant solution was added to wash out and collect the LS nanotubes.

**UV-Vis-NIR absorption measurements.** UV-Vis-NIR absorption spectra of water, 1% Intralipid in water and acetone, the mouse cranial bone and scalp skin were measured by a Cary 6000i UV-Vis-NIR spectrophotometer, background-corrected for contribution from any solvent. The measured range was 600-1800 nm. For water and 1% Intralipid in acetone, since there was no scattering of light by the samples, only absorption spectra were measured. For 1% Intralipid in water and the mouse tissue samples, since the UV-Vis-NIR spectrometer only collected light at the incident angle of the beam, the scattered incident light from the sample would not be collected and would thus add to the absorption spectra. Therefore the obtained absorption spectrum for any scattering sample should really be extinction spectra, which can be considered as a combination of the absorption spectrum and the scattering spectrum.

**NIR fluorescence spectroscopy of the SWNT-IRDye800 conjugate.** NIR fluorescence spectrum was taken using a home-built NIR spectroscopy setup in the 850-1650 nm region. The excitation light was provided by an 808-nm diode laser (RMPC lasers) at an output power of 160 mW and filtered by an 850-nm short-pass filter (Thorlabs), a 1000-nm short-pass filter (Thorlabs), an 1100-nm short-pass filter (Omega) and a 1300-nm short-pass filter (Omega). The excitation light was allowed to pass through the solution sample of SWNT-IRDye800 conjugate in a 1 mm path cuvette (Starna Cells) and the emission was collected in the transmission geometry. The excitation light was rejected using an 850-nm long-pass filter (Thorlabs) so that the fluorescence of both IRDye800 and SWNTs could be collected in the 850-1650 nm wavelength range. The emitted fluorescence from the solution sample was directed into a spectrometer (Acton SP2300i) equipped with a liquid-nitrogen-cooled InGaAs linear array detector (Princeton OMA-V). The emission spectrum was corrected after data acquisition to account for the laser excitation bleed-through, the sensitivity profile of the detector and extinction feature of the filter using the MATLAB software.

**Mouse handling, surgery and injection.** C57Bl/6 mice were obtained from Taconic Farms. All animal studies were approved by Stanford University's Administrative Panel on Laboratory Animal Care. Induction of middle cerebral artery occlusion (MCAO) and cerebral hypoperfusion was performed according to a previous study with minor modifications.[49] Briefly, mice were anesthetized using 30% $O_2$, 70% $N_2O$, and 2% isoflurane before surgery. For induction of MCAO, an incision was made on the left common carotid artery (CCA) and a silicon-coated nylon filament (Doccol Co., CA) was introduced into the left CCA and threaded forward into internal carotid artery (ICA) until the tip occludes the origin of the middle cerebral artery (MCA). For induction of cerebral hypoperfusion, only the left external carotid artery (ECA) and common carotid artery (CCA) were ligated. Control, unsurgerized mice ($n$=3), mice with induced MCAO ($n$=4) and mice with cerebral hypoperfusion ($n$=4) were used in the study. The hair over the scalp skin was removed using Nair before tail-vein injection and imaging. All mice were initially anesthesized before imaging in a knockdown box with 2 L min$^{-1}$ $O_2$ gas flow mixed with 3% Isoflurane. A nose cone delivered 1.5 L min$^{-1}$ $O_2$ gas and 3% Isoflurane throughout imaging. For comparative cerebrovascular imaging in NIR-I, NIR-II and NIR-IIa regions, a solution (200 µL) of 0.43 mg•mL$^{-1}$ (4.3 mg•kg$^{-1}$ body weight) SWNT-IRDye800 conjugates was injected into a mouse intravenously. For dynamic imaging in the NIR-IIa region, a solution (200 µL) of 0.43 mg•mL$^{-1}$ (4.3 mg•kg$^{-1}$ body weight) SWNT-DSPE-mPEG without IRDye800 was injected into a mouse intravenously. The maximum SWNT concentration in the blood circulation was ~4× lower than our previously found half maximal inhibitory concentration (IC50) of SWNTs to vascular endothelial cells.[22] The blood circulation half-time of DSPE-mPEG functionalized SWNTs was ~5 h,[20] and our previous studies had shown the lack of acute or long-term toxicity of such PEGylated SWNTs in vivo.[50-52] For the injection of nanotube solution, a 28 gauge syringe needle was inserted into the tail vein, allowing for bolus injection during the first frames of imaging. For steady-state imaging of the mouse head, injection was usually done ~5 min before the mouse was transferred to the imaging stage and imaged. For dynamic imaging of the mouse head, injection was done in the dark and the InGaAs camera started recording images continuously immediately after the nanotube solution was injected into the tail vein.

**NIR fluorescence imaging of phantom and mouse brain in different sub-regions.** A liquid-nitrogen-cooled, 320 × 256 pixel two-dimensional InGaAs array (Princeton Instruments) was used to take images in all sub-regions of NIR including the NIR-I, the NIR-II and the NIR-IIa regions. The excitation light was provided by an 808-nm diode laser (RMPC lasers) coupled to a collimator with a focal length of 4.5 mm (Thorlabs). The excitation light was filtered by an 850-nm short-pass filter and a 1000-nm short-pass filter (Thorlabs) before reaching the sample on the imaging plane. The excitation power density at the imaging plane was 140 mW cm$^{-2}$, significantly lower than the reported safe exposure limit of 329 mW cm$^{-2}$ at 808 nm.[53] The emitted fluorescence was allowed to pass through different filter sets to ensure the NIR images taken in different sub-regions. For the NIR-I region, an 850-nm long-pass filter and a 900-nm short-pass filter (Thorlabs) were used the confine the NIR-I region of 850-900 nm. For the NIR-II region, a 910-nm long-pass filter and a 1000-nm long-pass filter (Thorlabs) were used to confine the NIR-II region of 1000-1700 nm. The upper bound at 1700 nm was determined by the sensitivity profile of the InGaAs detector. For the NIR-IIa region, a 1000-nm long-pass filter, a 1300-nm long-pass filter (Thorlabs) and a 1400-nm short-pass filter (Edmund Optics) were used to confine the NIR-II region of 1300-1400 nm. A lens pair consisting of two achromats (200 mm and 75 mm, Thorlabs) was used to focus the image onto the detector with a field of view of 25 mm × 20 mm, which covered a segment of the SWNT-IRDye800-filled capillary tube in the phantom imaging case and only the head and the neck area of the mouse in the *in vivo* brain imaging case. A higher magnification was also achieved by using two other NIR achromats (150 mm and 200 mm, Thorlabs) to zoom into a smaller region of the brain. Phantom NIR fluorescence images were flat-field corrected to compensate for the non-uniformity of the excitation laser beam.

**High resolution microscopic imaging of cerebral vessels in the NIR-IIa region.** High-magnification intravital imaging of cerebral vessels was carried out in epifluorescence mode with an 808-nm diode laser (RMPC lasers, 160 mW) as the excitation source and two objective lens (4x and 10x, Bausch & Lomb) for microscopic imaging. The mouse with scalp hair removed was intravenously injected with a solution (200 μL) of 0.22 mg mL$^{-1}$ (2.2 mg kg$^{-1}$ body weight) separated LS nanotubes and placed in a home-made stereotactic platform fixed on a motorized 3D translational stage (Newport) that allowed for the digital position adjustment and readout of the mouse relative to the objective. The stereotactic stage had two posts fixed on two dovetail linear translational stages (Thorlabs) allowing for fine adjustment to fix the motion of the mouse head. The resulting NIR photoluminescence was collected using the same 2D InGaAs camera as aforementioned. The emitted fluorescence was filtered through a 1000-nm long-pass filter, a 1300-nm long-pass filter (Thorlabs) and a 1400-nm short-pass filter (Edmund Optics) to ensure only photons in the 1300–1400 nm NIR-IIa region were collected. Exposure times of 0.3–1 s were used for best signal-to-noise ratio. The depth of vessels was determined by first focusing the imaging plane onto the surface of the scalp skin, setting this as zero depth, and recording the axial distance the platform had travelled relative to the objective from the zero depth to reach a vascular image, based on the digital readout of the 3D translational stage. Since the objective was immersed in air, the raw axial movement of the objective was corrected to account for the refractive index difference of scalp skin ($n$=1.38),[54] cranial bone ($n$=1.56)[55] and brain tissue ($n$=1.35)[56] and to obtain the actual imaging depth inside the brain based on Snell's law.[8]

**Dynamic cerebral vascular imaging in the NIR-IIa window.** The dynamic imaging setup was the same as the aforementioned NIR-IIa fluorescence brain imaging with lower magnification that covered the entire mouse head. The InGaAs camera was set to expose continuously, and fluorescence images in the NIR-IIa window were acquired with LabVIEW software. The exposure time for each image acquisition was 100 ms. There was also an 87.5-ms overhead time in the readout, leading to a total time of 187.5 ms between consecutive frames and a frame rate of 5.3 frames s$^{-1}$ for the video. Depending on the orientation of the mouse head, the time-course images were rotated to make the head upright in the field of view using the MATLAB built-in function *imrotate*. This turned the left and right cerebral hemispheres symmetrical in the image and made it easier to select ROIs for cerebral blood perfusion measurement later. Then dynamic-contrast-enhanced images were obtained by loading 100 consecutive frames starting from the frame when signal first appeared in the brain, into an array using the MATLAB software, and the built-in *princomp* function was used to perform PCA.[20,22,57] The principal components featuring pixels showing up earlier in the video were automatically combined and used to represent arterial features while those featuring pixels showing up later in the video were combined and used to represent venous features based on the haemodynamic difference of arterial and venous flows.

**Brain blood perfusion measurement using NIR-IIa fluorescence.** In a typical procedure for brain blood perfusion measurement, consecutive NIR-IIa images taken from dynamic imaging were loaded into MATLAB software, and ROIs of both the surgerized left cerebral hemisphere and the untreated right cerebral hemisphere were selected in the lateral sulcus region. The NIR-IIa fluorescence intensity increase within each ROI was plotted as a function of time from 0 s to 3.94 s (22 frames) p.i. The plot featured a linear rising edge starting at ~2 s followed by a plateau region due to blood saturation of NIR-IIa contrast agent. Then the plot was normalized against the saturation level of the control cerebral hemisphere (i.e., the right hemisphere) and the linear rising edge after normalization was fitted into a line with its slope proportional to blood velocity as proved in previous publications.[22,58] By such analysis two slope values were obtained for each mouse, one derived from the control right cerebral hemisphere, and the other derived from the surgerized left cerebral hemisphere. Then the slope derived from the left hemisphere was normalized against that from right hemisphere to obtain the relative perfusion in the surgerized brain tissue to reveal the degree of occlusion of blood perfusion:

$$Relative\ Perfusion = \frac{slope\ of\ increase\ in\ the\ left\ hemisphere\ (surgerized)}{slope\ of\ increase\ in\ the\ right\ hemisphere\ (control)}$$

This measurement was repeated for 3-4 mice in each group to obtain statistically significant data.

**Vessel width analysis for NIR-IIa images.** To analyze the widths of blood vessels for *in vivo* cerebral vascular imaging, all images were loaded into the ImageJ software, and a line was drawn perpendicular to a linear feature of interest (i.e., a blood vessel). Then the NIR-IIa intensity values on this line were extracted and plotted against their physical locations along this line. The blood vessels intersected by the line were represented as peaks in the intensity profile and each peak was fitted into a Gaussian function using the Origin software.

**Laser Doppler flowmetry for cerebral blood perfusion measurement.** Mice were anesthetized with 1.5% isoflurane (1.5%), $O_2$ (30%) and $N_2O$ (70%) mixture. Body temperature was maintained at 36-37 ℃. A flexible fiberoptic probe was affixed to the skull over the MCA (2 mm posterior and 6.5 mm lateral to bregma) after the scalp skin was removed for cerebral blood flow (CBF) measurements by laser Doppler flowmetry (LDF). Baseline CBF values were measured before vascular intervention and considered to be 100% flow.

## Acknowledgments

This study was supported by grants from the National Cancer Institute of US National Institute of Health to H. D. (5R01CA135109-02), an American Heart Association Innovative Science Award and NIH 1R01NS064517 to C. J. K., and a William S. Johnson Fellowship to G. H. We acknowledge helpful discussions with Prof. John M. Pauly, Dr. Daryl Wong and Dr. Tao Zhang.

## Author contributions

H.D., C.J.K., G.H., S.D., and J.C. conceived and designed the experiments. G.H., S.D., J.C., A.L.A., C.C., B.Z., S.Z. and D.N.A. performed the experiments. G.H., S.D., J.C., A.L.A., C.C., B.Z., S.Z., D.N.A., P.L.H., K.I.A., C.J.K. and H.D. analyzed the data and wrote the manuscript. All authors discussed the results and commented on the manuscript.

# Figures

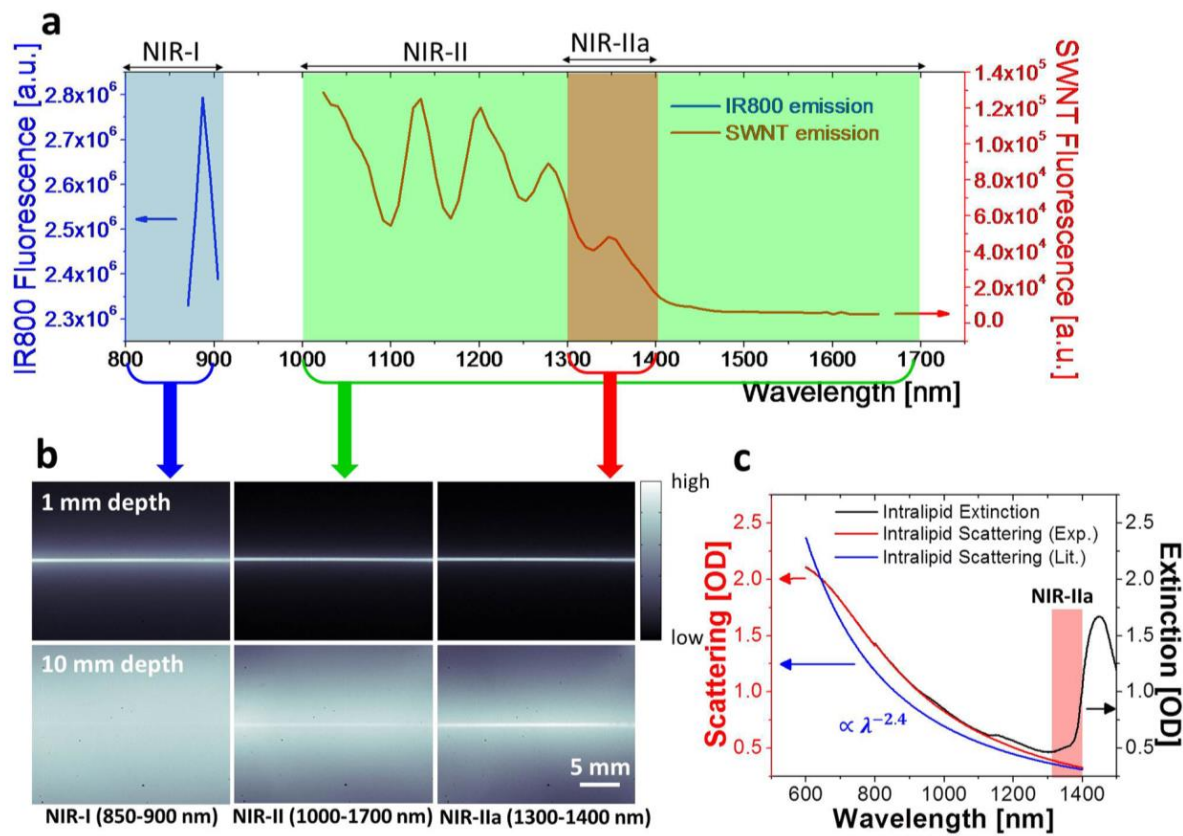

**Figure 1** Imaging in various NIR sub-regions. (**a**) A fluorescence emission spectrum of SWNT-IRDye800 conjugate in the range of 850-1650 nm under the excitation of an 808-nm laser. The emission spectra of IRDye800 and SWNT are plotted under different *y* axis scales to accommodate both into the same graph, due to the much higher fluorescence intensity of IRDye800 than SWNTs. (**b**) NIR fluorescence images of a capillary tube filled with SWNT-IRDye800 solution immersed at depths of 1 mm (top) and 10 mm (bottom) in 1% Intralipid recorded in NIR-I, NIR-II and NIR-IIa regions respectively. (**c**) The extinction spectrum (black curve) and scattering spectrum (red curve, measured by subtracting water and Intralipid absorptions from the extinction spectrum, see **Fig. S2**) of 1% Intralipid in water with path length of 1 mm measured by UV-Vis-NIR spectrometer, along with reduced scattering coefficient profile (blue, converted to the unit of OD with base-10 common log) of 1% Intralipid derived from literature.[33]

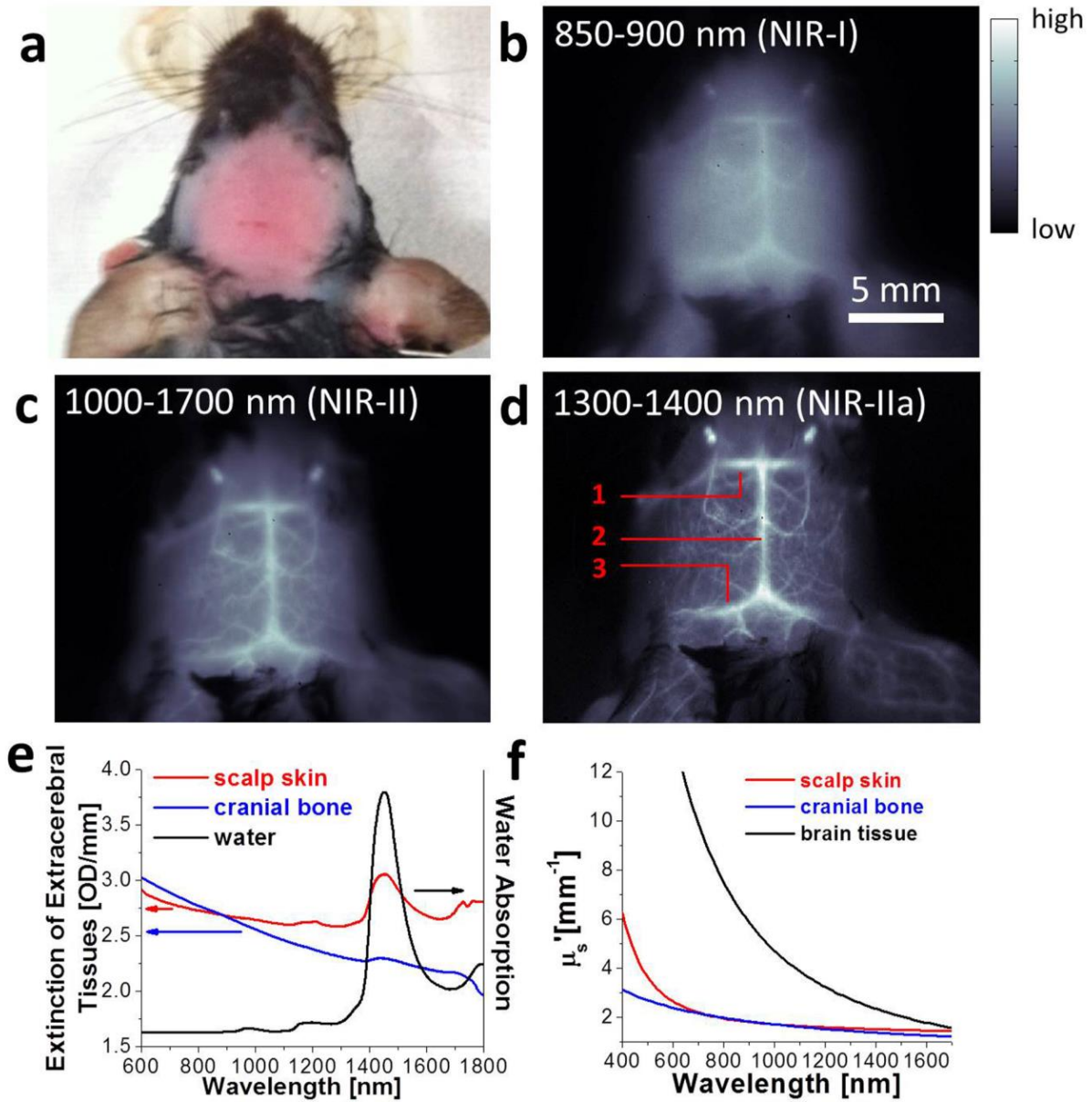

**Figure 2** *In vivo* mouse brain imaging with SWNT-IRDye800 in different NIR sub-regions. (**a**) A C57Bl/6 mouse head with hair removed. (**b-d**) Fluorescence images of the same mouse head in the NIR-I, NIR-II and NIR-IIa regions. The inferior cerebral vein, the superior sagittal sinus and transverse sinus are labeled 1, 2 and 3 in **d**, respectively. (**e**) Extinction spectra of scalp (red) and skull (blue) along with the water absorption spectrum (black). (**f**) Reduced scattering coefficients $\mu_s'$ of scalp skin (red), cranial bone (blue) and brain tissue (black) plotted against wavelength, based on the previously reported scattering properties for these tissues.[8,35,36]

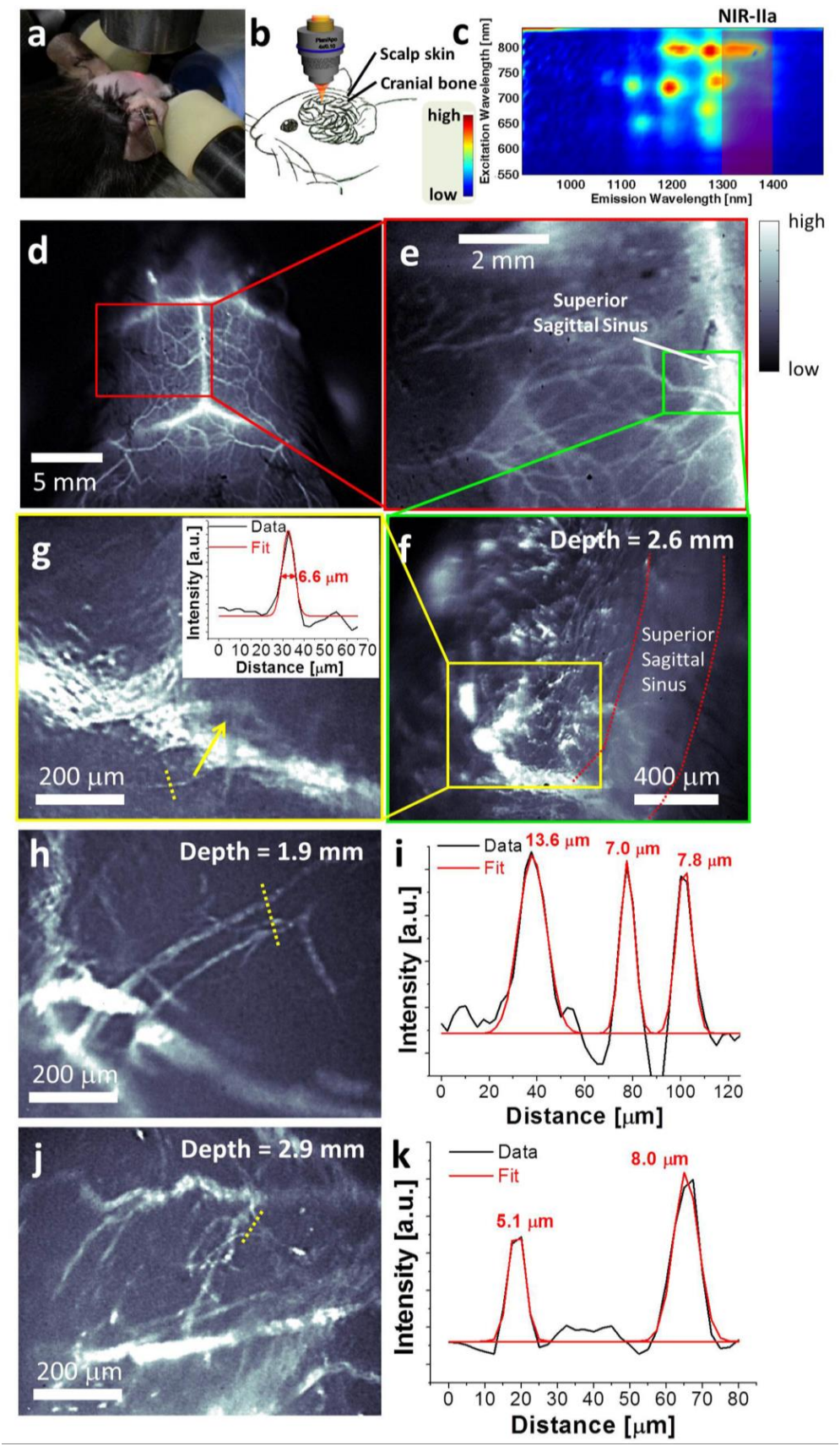

**Figure 3** Non-invasive, high-resolution NIR-IIa fluorescence imaging of mouse brain vasculature. (**a**) A photo showing the stereotactic microscopic imaging setup, where a red laser is used for alignment and shows the beam location. (**b**) A schematic showing the penetration of NIR-IIa fluorescence through brain tissue, skull and the scalp. (**c**) A photoluminescence versus excitation (PLE) spectrum of LS nanotubes in an aqueous solution. The 1.3-1.4 μm NIR-IIa region is shaded red. (**d**) A low-magnification cerebral vascular image taken with a field of view of 25 mm ×20 mm. (**e**) A cerebral vascular image of the same mouse head zoomed into the left cerebral hemisphere, with a field of view of 8 mm ×6.4 mm. (**f**) A cerebral vascular image of the same mouse head taken using a microscope objective, with a field of view of 1.7 mm ×1.4 mm. The depth of these in-focus vascular features was determined as 2.6 mm. (**g**) A zoomed-in image of a sub-region in **f** taken by a higher magnification objective, with a field of view of 0.80 mm × 0.64 mm. The inset shows the cross-sectional intensity profile (black) and Gaussian fit (red) along the yellow-dashed bar. (**h-k**) Two other high resolution cerebral vascular images with a field of view of 0.80 mm ×0.64 mm taken on another mouse (**h&j**), and their cross-sectional fluorescence intensity profiles (black) and Gaussian fit (red) along the yellow-dashed bars (**i&k**).

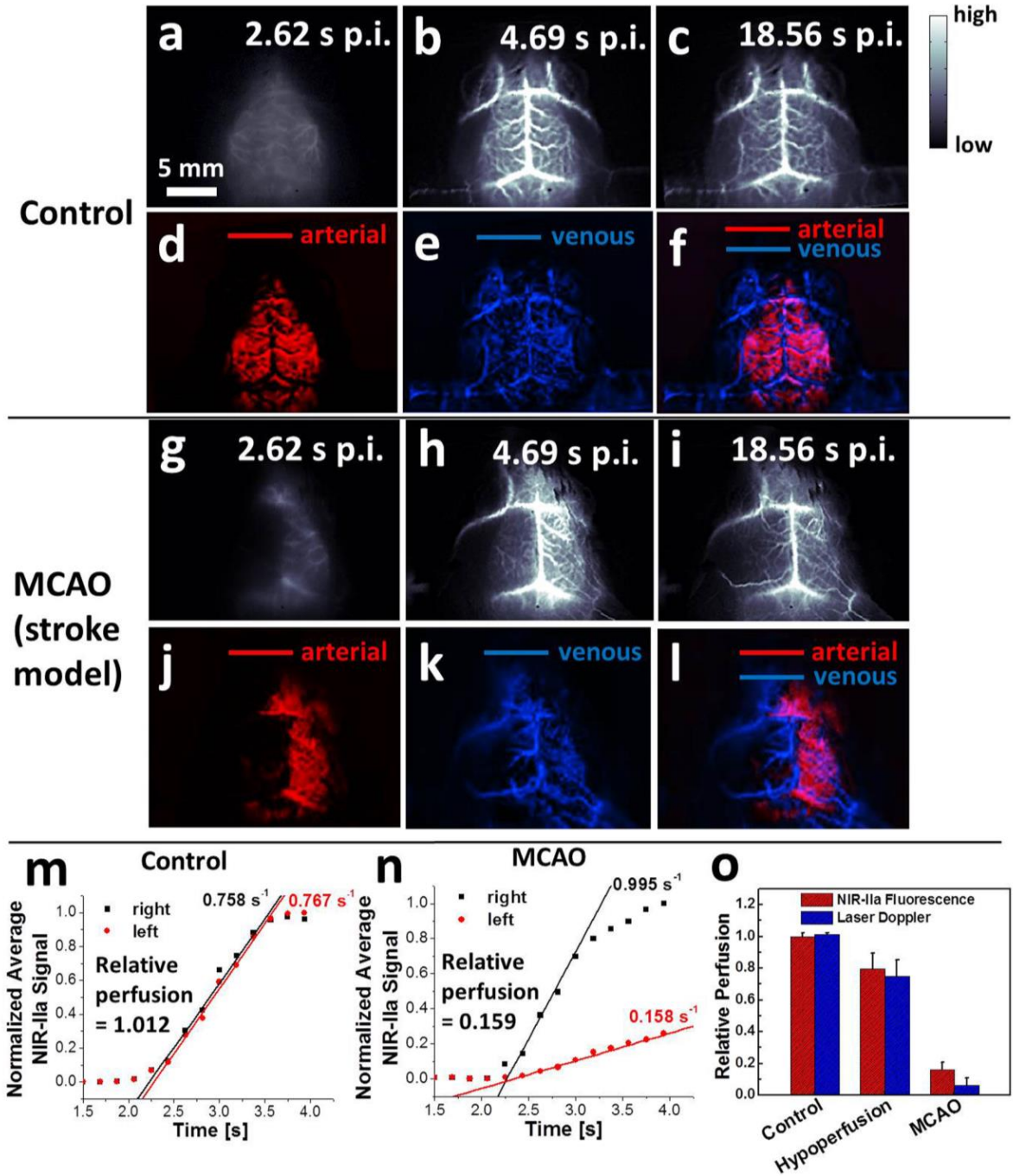

**Figure 4** Dynamic NIR-IIa fluorescence imaging of mouse cerebral vasculature. (**a-c**) Time course NIR-IIa images of a control, healthy mouse (Mouse C1). (**d-f**) PCA overlaid images showing arterial (red) and venous (blue) vessels of Mouse C1. (**g-i**) Time course NIR-IIa images of a mouse with MCAO (Mouse M1). (**j-l**) PCA overlaid images showing arterial (red) and venous (blue) vessels of Mouse M1. (**m-n**) Normalized NIR-IIa signal in the left (red) and right (black) cerebral hemispheres of Mouse C1 (**m**) and M1 (**n**) versus time. (**o**) Average blood perfusion of the left cerebral hemisphere of control group (*n*=3), MCAO group (*n*=4) and cerebral hypoperfusion group (*n*=4), measured by NIR-II method (red) and laser Doppler blood spectroscopy (blue). Errors bars reflect the standard deviation of each group.

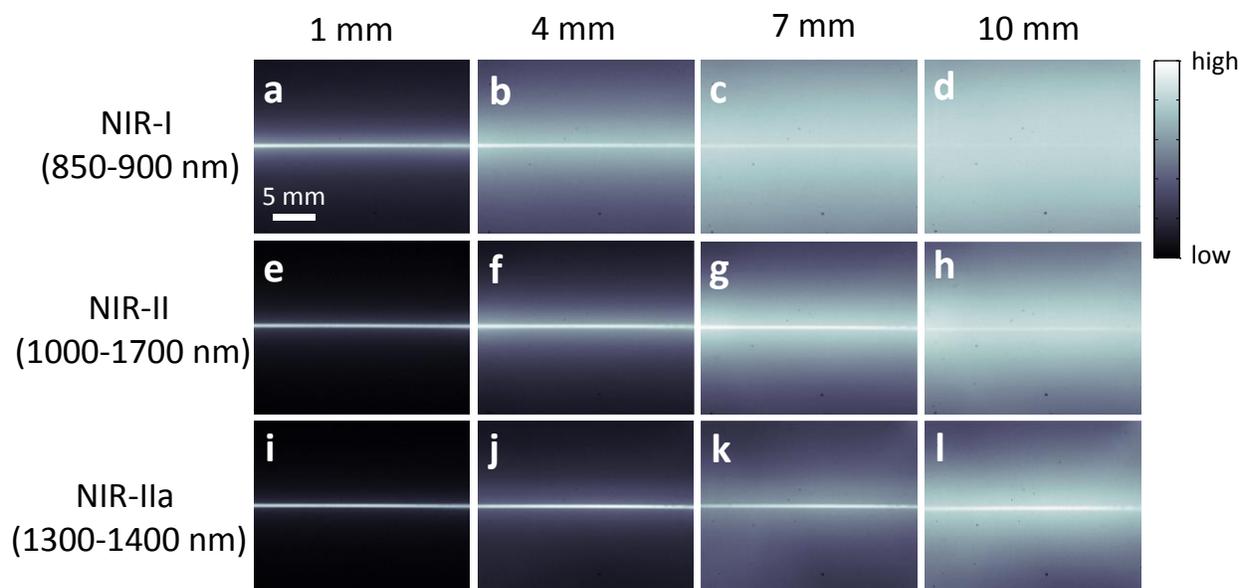

**Figure S1** NIR fluorescence images of a capillary tube filled with SWNT-IRDye800 solution immersed at depths of 1 mm (**a**, **e**, **i**), 4 mm (**b**, **f**, **j**), 7 mm (**c**, **g**, **k**) and 10 mm (**d**, **h**, **l**) in 1% Intralipid recorded in NIR-I (**a-d**), NIR-II (**e-h**) and NIR-IIa (**i-l**) regions respectively.

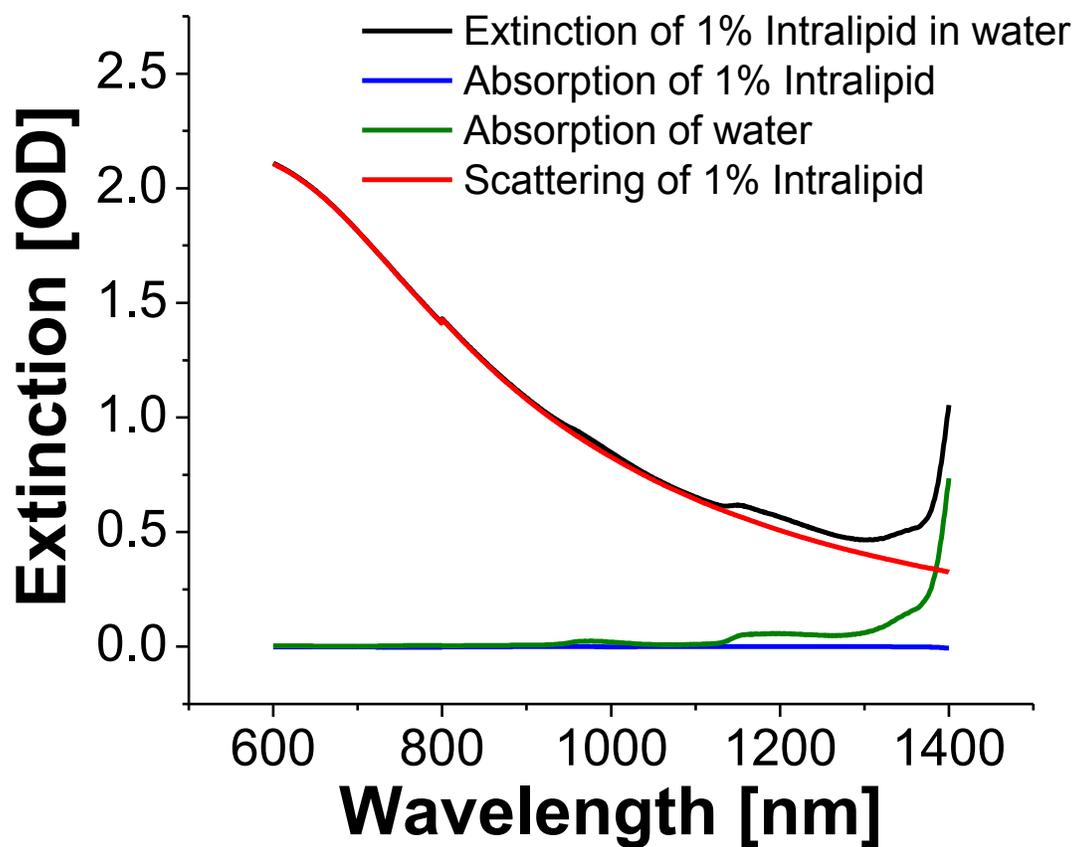

**Figure S2** Determination of the scattering spectrum of 1% Intralipid in water. The scattering spectrum (red curve) was obtained by subtracting the absorption of water (green curve) and absorption of 1% Intralipid in an organic solvent (blue curve; acetone forms a non-scattering solution for Intralipid) from the as-measured extinction spectrum of 1% Intralipid aqueous solution (black curve).

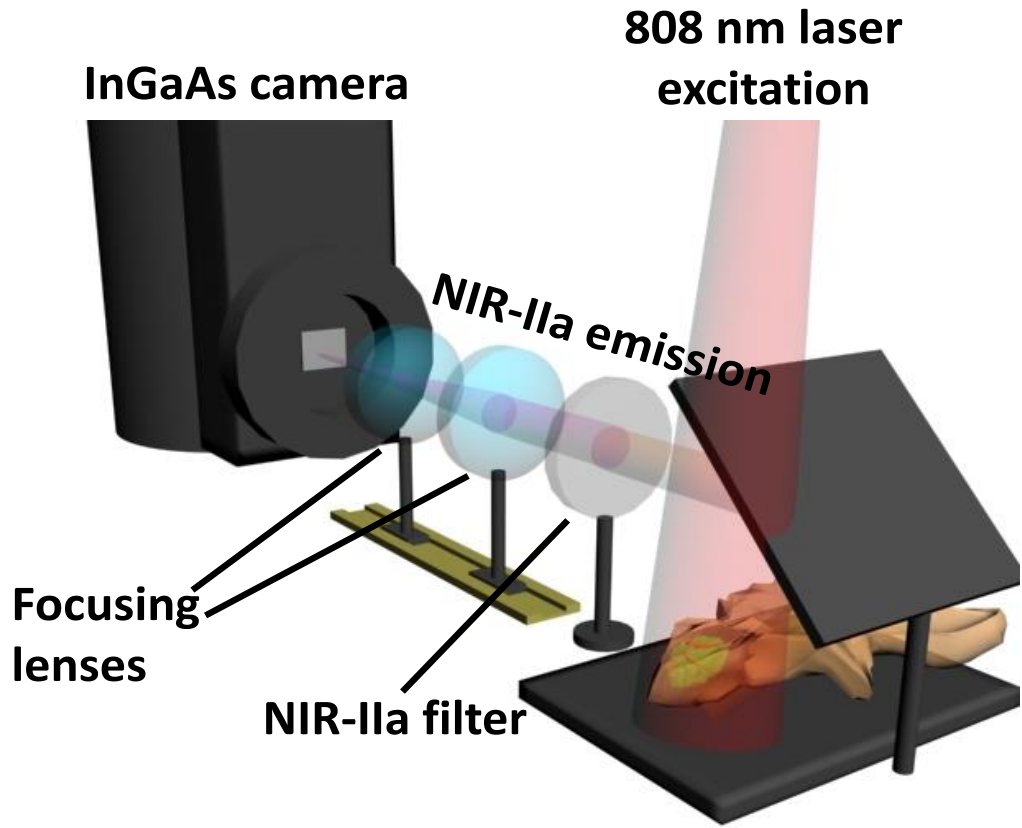

**Figure S3** A schematic of the NIR-IIa fluorescence imaging rig for non-invasive through-scalp and through-skull brain vascular imaging.

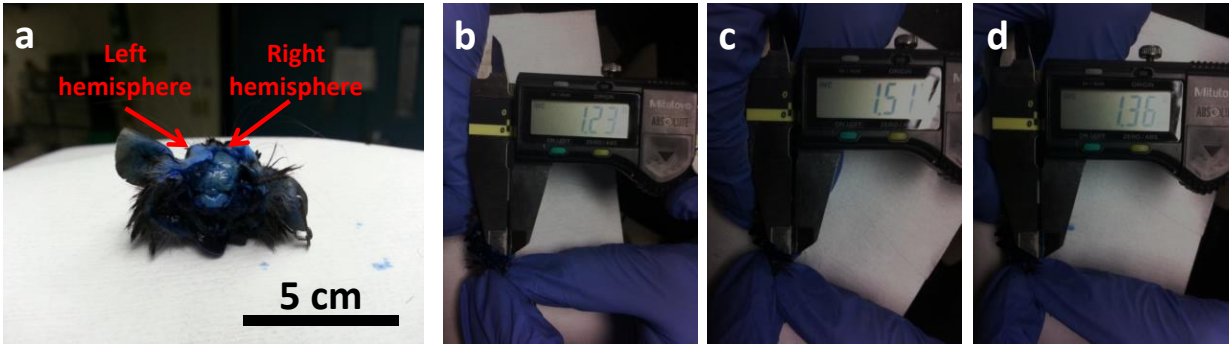

**Figure S4** Physical measurement of the total thickness of scalp skin, cranial bone and the meninges of a C57Bl/6 mouse. (**a**) A digital camera photo of a dissected mouse head. The scalp, skull and meninges over the right cerebral hemisphere were completed removed while those over the left hemisphere remained intact. Note that the mouse was perfused with 5% Evans blue before sacrifice to render the vessels blue in the brain. (**b-d**) A repeated measure of the total thickness of scalp skin, cranial bone and the meninges by digital calipers at three different locations. The average thickness was 1.37 mm, setting the lower bound of cerebral vessel depths in our imaging experiments. Therefore the cortical vessels located at the surface of the brain, including the inferior cerebral vein, the superior sagittal sinus and the transverse sinus, should have depths within the range of 1-2 mm underneath the scalp skin.

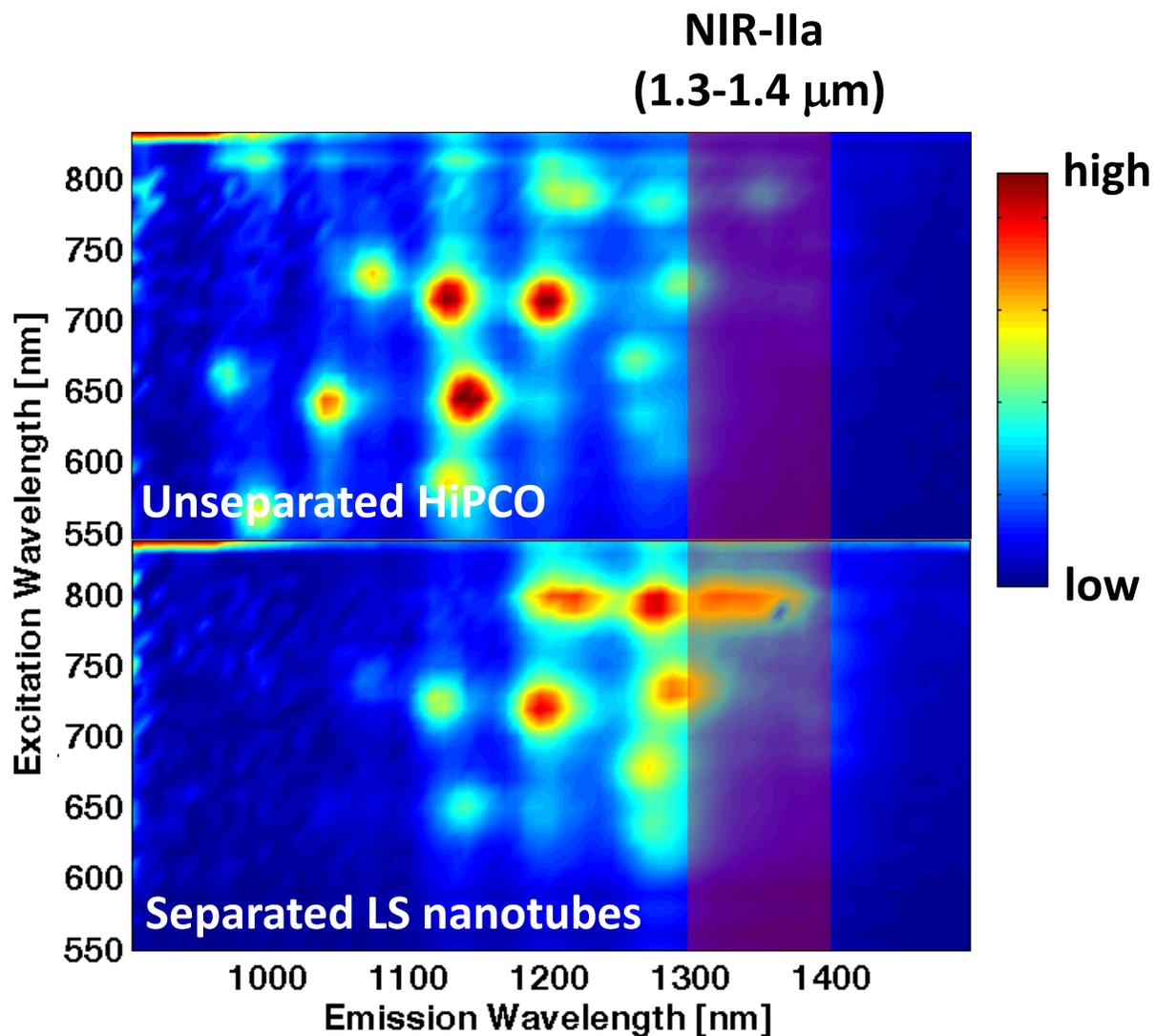

**Figure S5** Photoluminescence versus excitation (PLE) spectra of unseparated HiPCO SWNTs (top) and separated LS nanotubes (bottom) in aqueous solutions. The 1.3-1.4 μm NIR-IIa region is shaded red in both spectra to highlight the difference of fluorescence emission of the two samples in this region (enhanced emission in the region for separated LS nanotubes). Note that metallic nanotubes with no fluorescence were mostly removed in the separated LS sample.

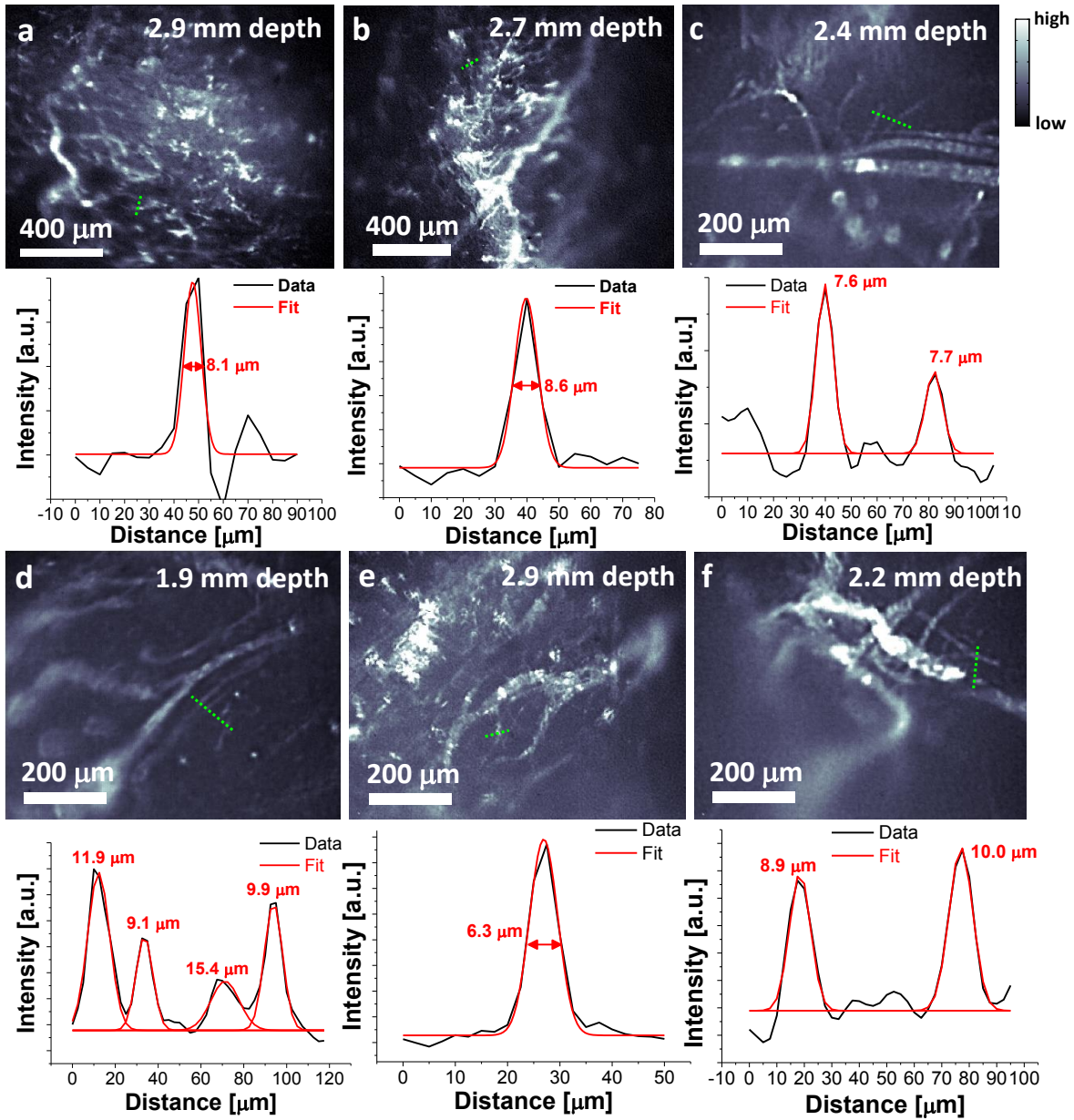

**Figure S6** More microscopic images of brain vessels recorded in the NIR-IIa region at depths in the range of 1-3 mm underneath the surface of the scalp skin of several mice. Images **a&b** were taken under a 4x microscopic objective with a field of view of 1.7 mm × 1.4 mm, while images **c-f** were taken under a 10x microscopic objective with a field of view of 800 μm × 640 μm. Cross-sectional fluorescence intensity profiles (black) along the green-dashed bars in the images are shown under each corresponding image, and Gaussian fitted peaks to the profiles are shown as the red curves, featuring capillary vessel widths in the range of 6-16 μm.

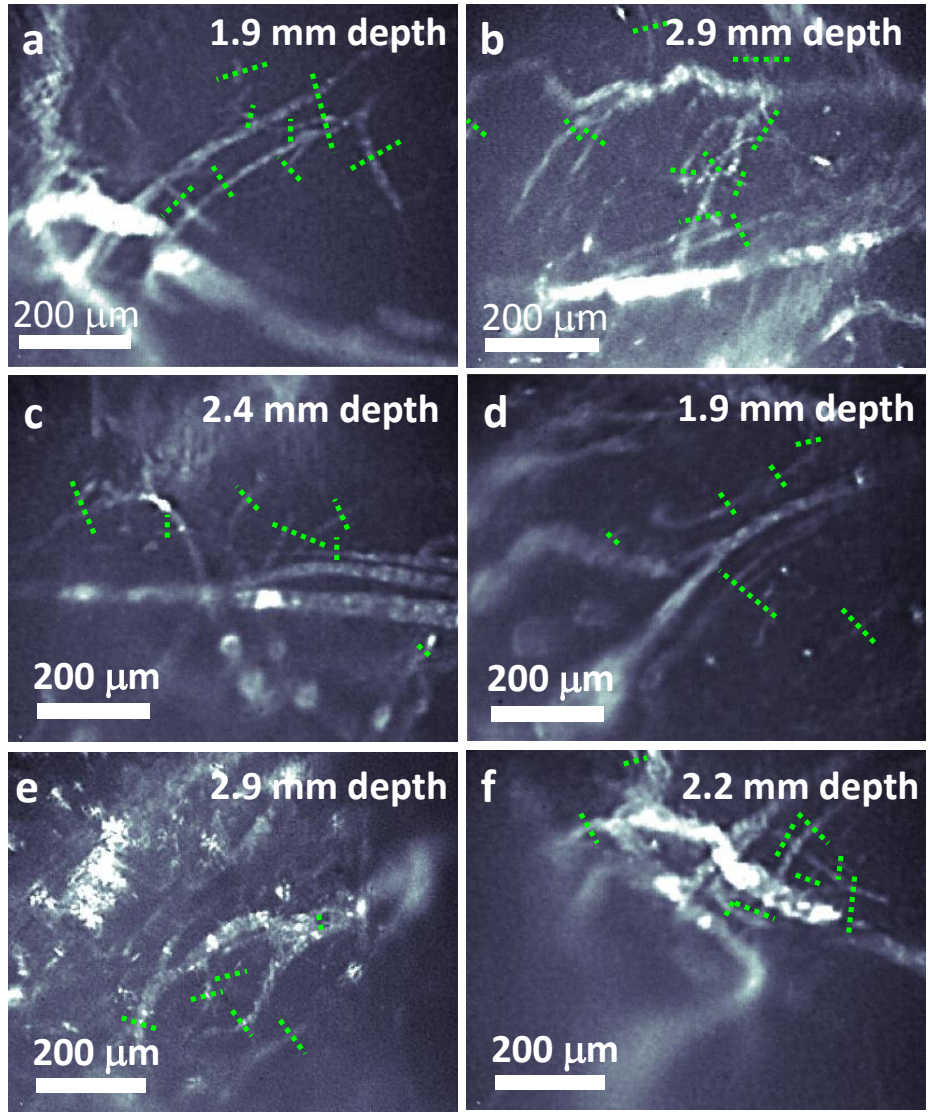

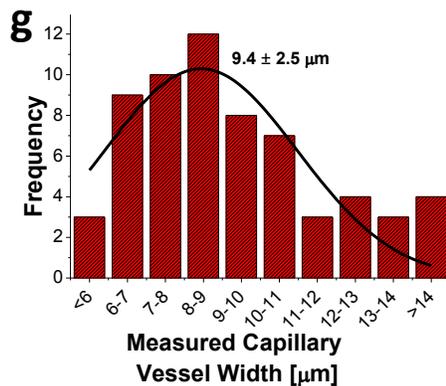

**Figure S7 Statistical analysis of capillary vessel widths in NIR-IIa fluorescence images.** (**a-f**) Microscopic cerebral vascular images with the analyzed capillary vessels intersected with green dashed bars. (**g**) A histogram showing the distribution of measured capillary vessel widths in the range of 5-15 μm, with a mean vessel width of 9.4 μm and a standard deviation of 2.5 μm.

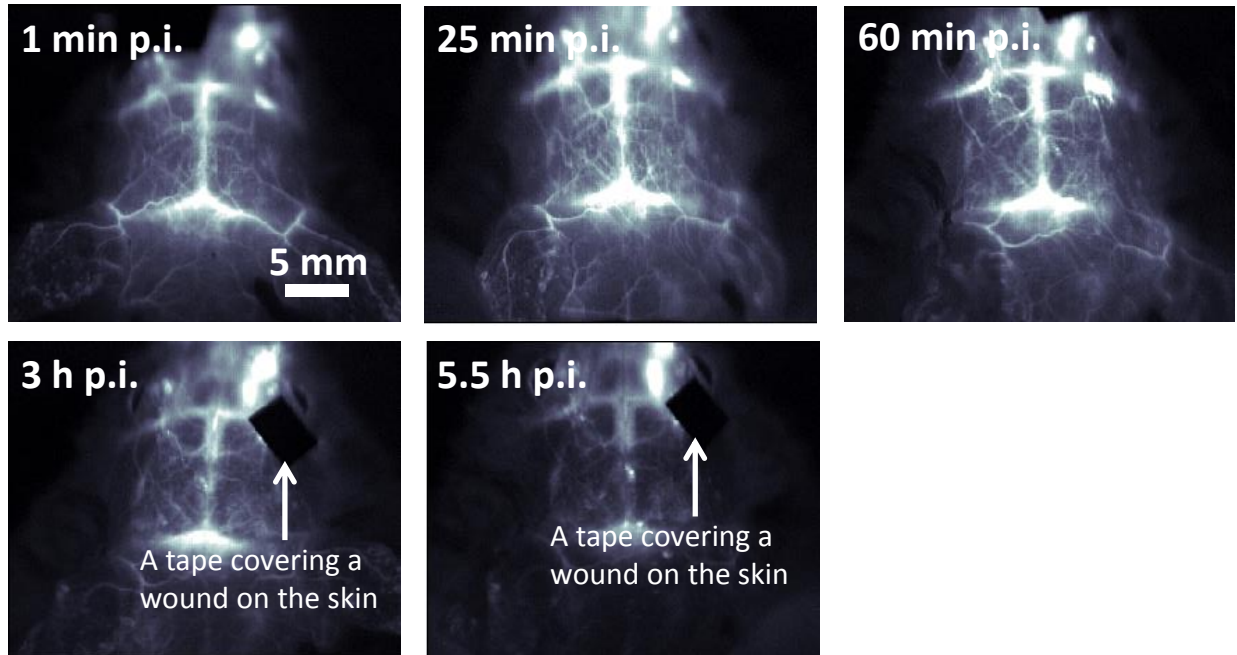

**Figure S8** NIR-IIa fluorescence images taken at different time points post injection (p.i.), between which we returned the mouse to its cage for it to recover from anesthesia. These time-course images show a time window of ~3 h p.i. for static brain vessel imaging. Note that in the last two images a tape was applied on part of the scalp skin to cover a wound possibly caused by the bite from another mouse in the same cage.

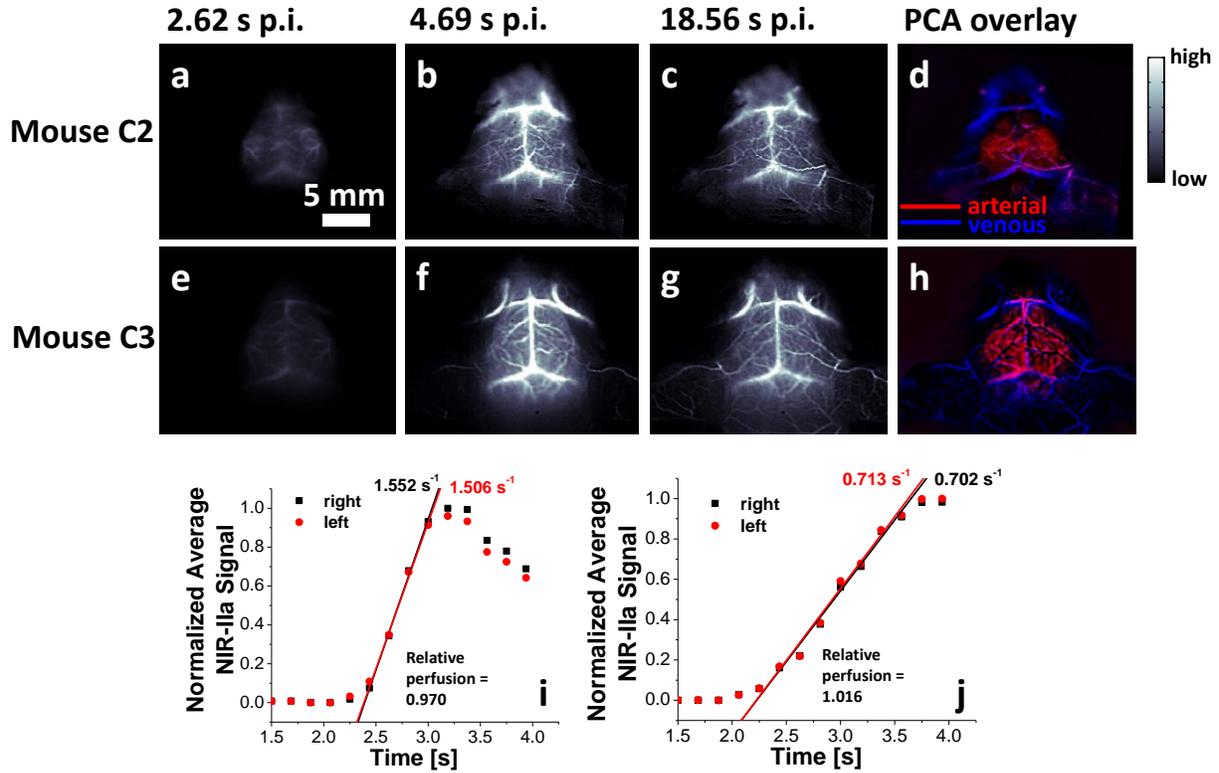

**Figure S9** Dynamic NIR-IIa fluorescence imaging of mouse cerebral vasculature on a group of control mice (Mouse C2 & C3) at a frame rate of 5.3 frames•s⁻¹. (**a-c**) Time course NIR-IIa fluorescence images showing blood flow in the cerebral vasculature of Mouse C2. (**d**) PCA overlaid image based on the first 100 frames (18.75 s post injection) of Mouse C2. (**e-g**) Time course NIR-IIa fluorescence images showing blood flow in the cerebral vasculature of Mouse C3. (**h**) PCA overlaid image based on the first 100 frames (18.75 s post injection) of Mouse C3. (**i**) Normalized NIR-IIa signal in the left (red) and right (black) cerebral hemispheres of Mouse C2 versus time. (**j**) Normalized NIR-IIa signal in the left (red) and right (black) cerebral hemispheres of Mouse C3 versus time.

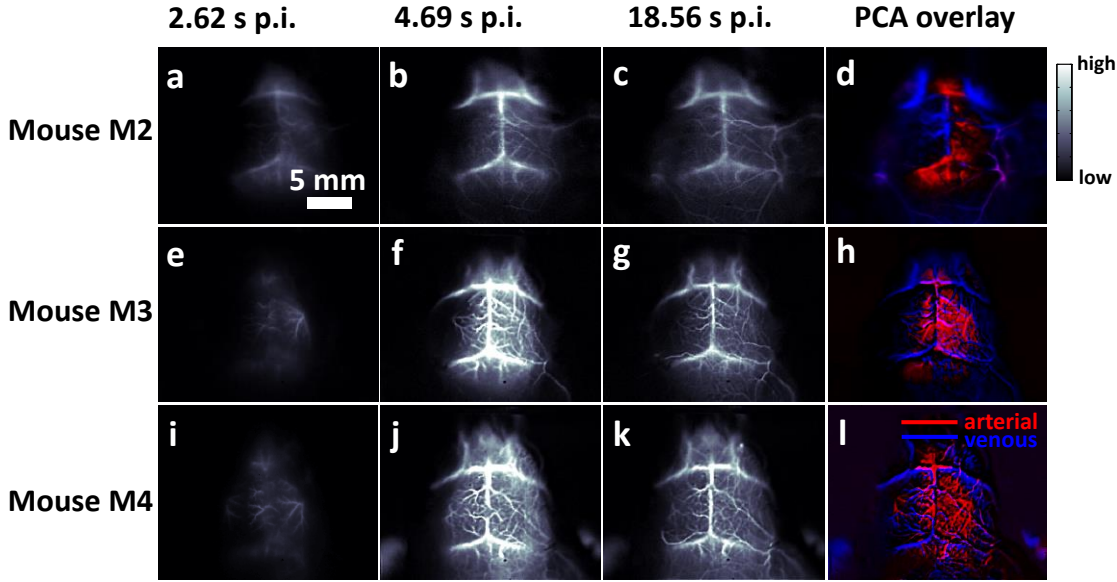

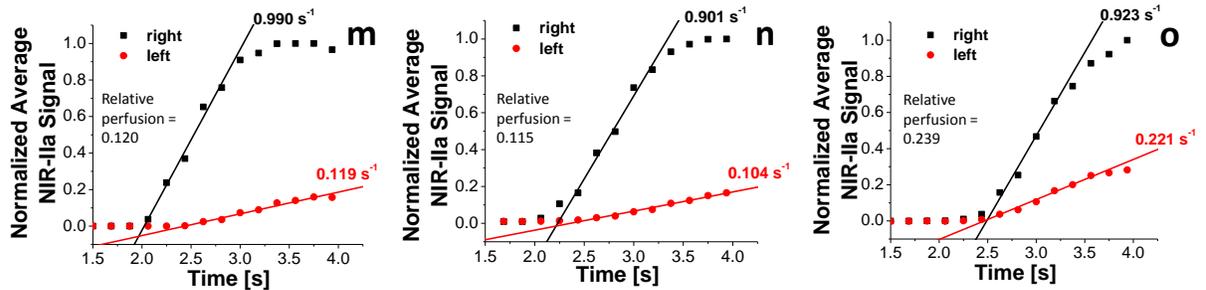

**Figure S10** Dynamic NIR-IIa fluorescence imaging of mouse cerebral vasculature on a group of MCAO mice (Mouse M2-4) at a frame rate of 5.3 frames•s$^{-1}$. (**a-c**) Time course NIR-IIa fluorescence images showing blood flow in the cerebral vasculature of Mouse M2. (**d**) PCA overlaid image based on the first 100 frames (18.75 s post injection) of Mouse M2. (**e-g**) Time course NIR-IIa fluorescence images showing blood flow in the cerebral vasculature of Mouse M3. (**h**) PCA overlaid image based on the first 100 frames (18.75 s post injection) of Mouse M3. (**i-k**) Time course NIR-IIa fluorescence images showing blood flow in the cerebral vasculature of Mouse M4. (**l**) PCA overlaid image based on the first 100 frames (18.75 s post injection) of Mouse M4. (**m**) Normalized NIR-IIa signal in the left (red) and right (black) cerebral hemispheres of Mouse M2 versus time. (**n**) Normalized NIR-IIa signal in the left (red) and right (black) cerebral hemispheres of Mouse M3 versus time. (**o**) Normalized NIR-IIa signal in the left (red) and right (black) cerebral hemispheres of Mouse M4 versus time.

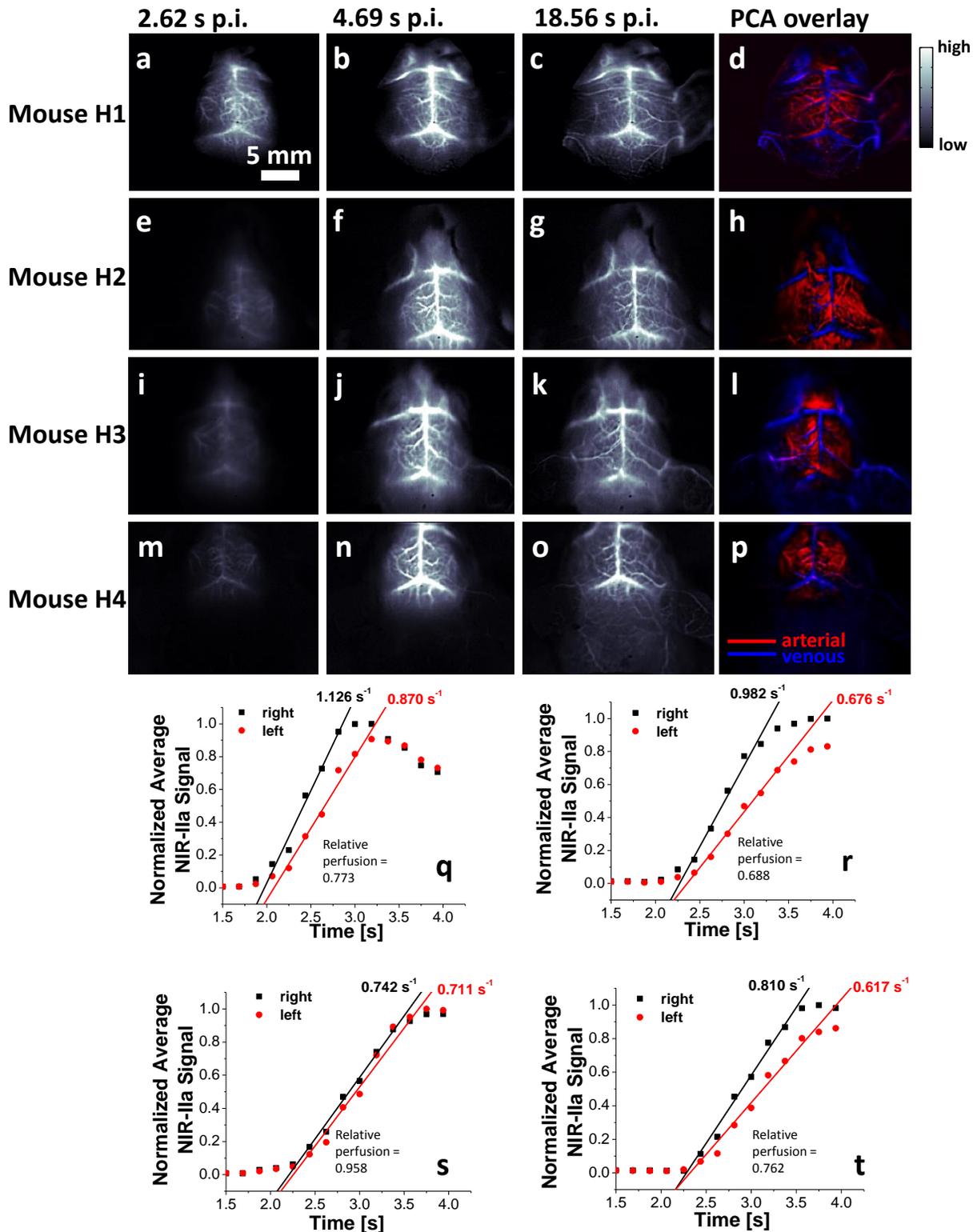

**Figure S11** Dynamic NIR-IIa fluorescence imaging of mouse cerebral vasculature on a group of mice with cerebral hypoperfusion (Mouse H1-4) at a frame rate of 5.3 frames•s⁻¹. (**a-c**) Time course NIR-IIa fluorescence images showing blood flow in the cerebral vasculature of Mouse H1. (**d**) PCA overlaid image based on the first 100 frames (18.75 s post injection) of Mouse H1.

(**e-g**) Time course NIR-IIa fluorescence images showing blood flow in the cerebral vasculature of Mouse H2. (**h**) PCA overlaid image based on the first 100 frames (18.75 s post injection) of Mouse H2. (**i-k**) Time course NIR-IIa fluorescence images showing blood flow in the cerebral vasculature of Mouse H3. (**l**) PCA overlaid image based on the first 100 frames (18.75 s post injection) of Mouse H3. (**m-o**) Time course NIR-IIa fluorescence images showing blood flow in the cerebral vasculature of Mouse H4. (**p**) PCA overlaid image based on the first 100 frames (18.75 s post injection) of Mouse H4. (**q**) Normalized NIR-IIa signal in the left (red) and right (black) cerebral hemispheres of Mouse H1 versus time. (**r**) Normalized NIR-IIa signal in the left (red) and right (black) cerebral hemispheres of Mouse H2 versus time. (**s**) Normalized NIR-IIa signal in the left (red) and right (black) cerebral hemispheres of Mouse H3 versus time. (**t**) Normalized NIR-IIa signal in the left (red) and right (black) cerebral hemispheres of Mouse H4 versus time.